\documentclass[a4paper,11pt]{article}
\usepackage{pos}

\title{On the threshold behaviour of heavy top production}

\author*[a]{Torbj\"orn Sj\"ostrand}

\affiliation[a]{Department of Physics, Lund University, \\
  Box 118, SE-221 00 Lund, Sweden}

\emailAdd{torbjorn.sjostrand@fysik.lu.se}

\abstract{The observation of an excess of $\t\tbar$ production in
the threshold region, by CMS and ATLAS, has been interpreted as a
toponium contribution, i.e. from below-threshold $\t\tbar$ virtual
states. The news here is the nontrivial experimental extraction of
such a signal, not its existence as such. Indeed, already 35+ years
ago an NRQCD Green's function approach was used to model the above- 
and below-threshold production of $\t\tbar$ pairs in $\p\p/\p\pbar$
collisions. The relevant cross section equations from that study are
now (re-)implemented in the \textsc{Pythia}~8 event generator. 
While the above-threshold part is straightforward, the physical
interpretation and modelling of below-threshold events is nontrivial,
and a final prescription is cross-checked against two simpler ones.
Cross sections and some event properties are presented.}

\FullConference{QCD at the Extremes (QCDEX2025)\\
1–5 Sept 2025\\
Online\\}

\newcommand{\e}{{\mathrm e}}
\newcommand{\g}{{\mathrm g}}
\newcommand{\p}{{\mathrm p}}
\newcommand{\q}{{\mathrm q}}
\renewcommand{\t}{{\mathrm t}}
\newcommand{\pbar}{\overline{\mathrm p}}
\newcommand{\qbar}{\overline{\mathrm q}}
\newcommand{\tbar}{\overline{\mathrm t}}

\begin{document}
\maketitle

This writeup is one of four dedicated to the study of the $\t\tbar$
threshold region, and notably the contribution from  
below-threshold $\t\tbar$ virtual states. The experimental
background, i.e.\ the recent CMS \cite{CMS:2025kzt} and ATLAS
\cite{ATLAS:2025top} observation of a cross section excess, is
covered by Baptiste Ravina. Two examples of recent theoretical
studies are presented by Benjamin Fuks \cite{Fuks:2024yjj} and
Sven-Olaf Moch \cite{Garzelli:2024uhe}. Here we revive what may be
the first study of the heavy-top threshold region in $\p\p/\p\pbar$
collisions, from 1989/90 \cite{Fadin:1989fd,Fadin:1990wx}, as a
follow-up of the earlier $\e^+\e^-$ study by Fadin and Khoze
\cite{Fadin:1987wz}. See also \cite{Fadin:1991zw}.

Relative to the Born-level $\t\tbar$ production cross section, a first
extension is to consider the Coulomb effects of gluon exchange between
the outgoing $\t$ and $\tbar$. Multiple gluon exchanges can be resummed
to an all-orders expression. This leads to an enhanced cross section
when the $\t\tbar$ pair is in a colour singlet state, and a reduced one 
when in a colour octet state. Based on relative colour factors we will 
assume that $\g\g \to \t\tbar$ is 2/7 singlet and 5/7 octet, while
$\q\qbar \to \t\tbar$ is all octet. Higher-order corrections,
e.g.\ from the emission of a gluon, could change these numbers.

The Coulomb expressions do not take into account the finite-top-width
effects, nor the below-threshold states. Instead NRQCD Green's function
expressions can be derived for the behaviour in the threshold region. 
These contain both continuum terms and an infinite (but rapidly 
convergent) sum over the virtual states, (pseudo)bound for the singlet
case. 

\begin{figure}
\includegraphics[width=0.615\textwidth]{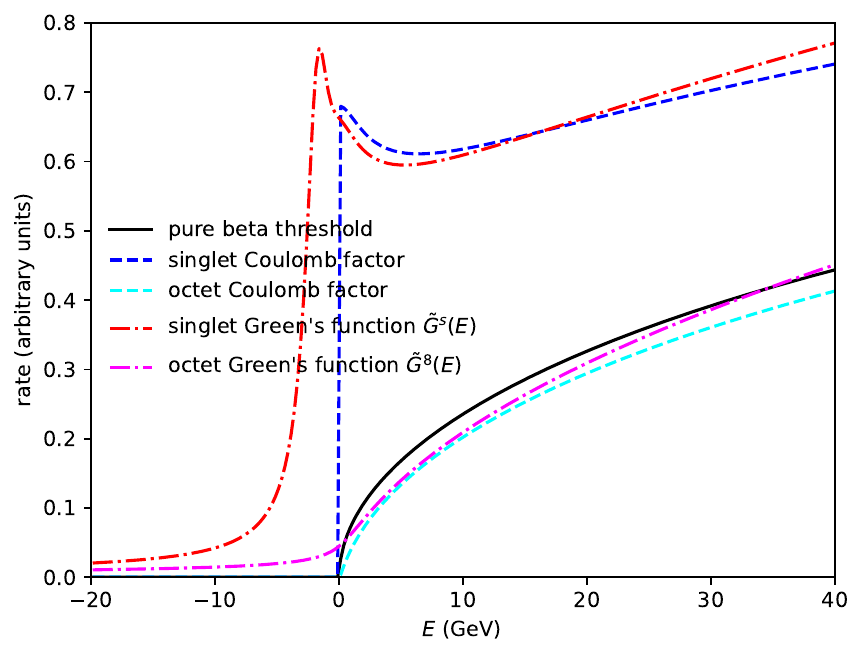}%
\includegraphics[width=0.385\textwidth]{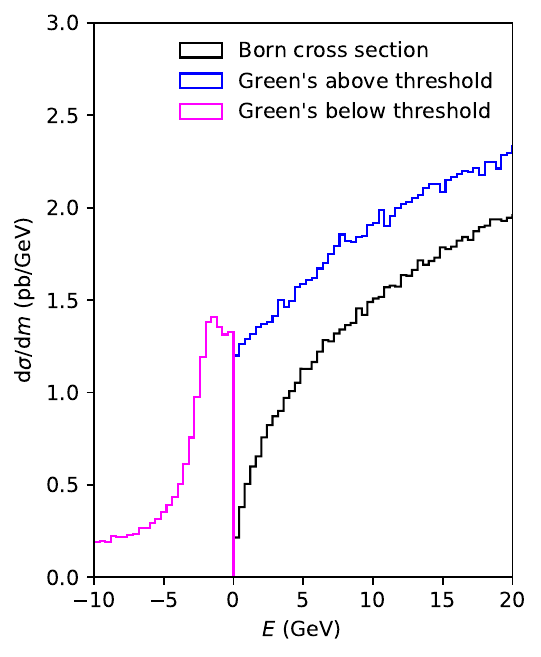}\\[-2mm]
\hspace*{0.31\textwidth}\textit{(a)}\hspace*{0.47\textwidth}\textit{(b)}
\caption{The threshold behavior.
\textit{(a)} Analytical expressions for singlet and octet cases,
$E = m_{\t\tbar} - 2m_{\t}$.
\mbox{\textit{(b)} Cross section in a four-mass scenario} simulation for 
the expected singlet/octet mix, $E = \hat{m} - m_{\t 1} - m_{\t 2}$.}
\label{fig:threshold}
\end{figure}

Newly encoded old formulae were compared with the 1990 results, for 
140 and 200 GeV top masses, and agreement was found. In
Fig.~\ref{fig:threshold}\textit{a} results are shown for a current
$m_{\t} = 172.5$ GeV value. Of note is that the Green's functions
diverge above the threshold region. In our numerical studies,
therefore a smooth transition to the more reliable Coulomb expression
is done between 10 and 20 GeV above the threshold. In a full event
generation setup, one may instead prefer to transition to an NLO
description \cite{Nason:1987xz}. Correspondingly, the below-threshold
contribution is damped to zero between $-10$ and $-20$ GeV.

The question now arises how to apply these equations in practice. 
For the above-threshold part the basic prescription is to generate
a $\t\tbar$ pair according to an overall weight
\begin{center}
BW($\t$) $\times$ BW($\tbar$) $\times$ PDF($x_1$) 
$\times$ PDF($x_2$) $\times$ ME $\times$ PS 
$\times$ ($\tilde{G}(E)/\beta_{\t}$)~,
\end{center}
where a standard accept/reject step is added to generate events with
unit weight. Here BW denotes the Breit-Wigner mass distributions,
PDF the parton distributions of the incoming gluons or quarks,
ME the (squared) matrix elements for $\g\g \to \t\tbar$ and
$\q\qbar \to \t\tbar$ respectively, and PS the phase space of the
production process. Note that the PS expression contains a factor
\begin{equation*}
\beta_{\t} = \sqrt{ \left( 1 - \frac{m_{\t 1}^2}{\hat{s}}
  - \frac{m_{\t 2}^2}{\hat{s}} \right) - 4 \, \frac{m_{\t 1}^2}{\hat{s}}
  \, \frac{m_{\t 2}^2}{\hat{s}} } ~,
\end{equation*}
where $m_{\t 1}$ and $m_{\t 2}$ are the two BW-selected masses, while
$m_{\t}$ is reserved for the on-shell value, and $\hat{s} = \hat{m}^2$ is
the squared invariant mass of the $\t\tbar$ system. In the final factor,
that takes us beyond the Born expression, the $\beta_{\t}$ is replaced
by the rescaled Green's  expression
$\tilde{G}(E) = (4\pi/m_{\t}^2)\, \mathrm{Im}\, G(E)$, where
$E = \hat{m} - m_{\t 1} - m_{\t 2}$. Note that the BW factors smear the 
threshold $E = 0$ away from being at a fixed $2 m_{\t} = 345$~GeV
to become an event-by-event number.

This approach does not work for $E < 0$, where formally phase space 
is vanishing. But a below-threshold state is subject to the weak decays
of the $\t$ and $\tbar$. From the final state a theorist's detector would
reconstruct some $m'_{\t 1}$ and $m'_{\t 2}$, where 
$m'_{\t 1} + m'_{\t 2} < m_{\t 1} + m_{\t 2}$
so that $E' = \hat{m} - m'_{\t 1} - m'_{\t 2} > 0$.
Therefore the $E < 0$ states are never detectable as such, but show up
as a deformation of the BW distributions, lowering the average mass.
We will strive towards such an implementation, via two simpler models
that will act as sanity checks.

The mirror solution is very simple: mirror the $E < 0$ contribution, 
so that events with $E > 0$ get a total weight  
$(\tilde{G}(E) + \tilde{G}(-E))/\beta_{\t}$. This is a 
robust Monte Carlo procedure that works for a poor experimental 
$\t\tbar$ mass resolution. The drop of the PDFs with $x$
gives too low an integrated cross section, however, and the BW 
distribution is not deformed. 

In the mass shift solution two separate runs are needed, one with
$\tilde{G}(E)/\beta_{\t}$ and another with $\tilde{G}(-E)/\beta_{\t}$, 
where the latter is with a shifted $m_{\t} = 172.5 \to 168.7$ GeV.
Thereby $\langle E \rangle = +3.82 \to -3.90$~GeV for the
$\tilde{G}(-E)$ part relative to the $m_{\t} = 172.5$ GeV threshold,
so that average PDF weights work out, but some kinematics
distributions are still off.

The preferred four-mass hybrid solution is more tricky. In it the
$m_{\t 1}$ and $m_{\t 2}$ masses are first picked. Phase space is then
sampled in the extended region $\hat{m} > m_{\t 1} + m_{\t 2} - 20$
GeV, i.e.\ as far below threshold as $\tilde{G}(E)$ is non-vanishing.
If the chosen $\hat{m}$ gives an $E > 0$ then the normal 
$\tilde{G}(E)/\beta_{\t}$ weight is applied. If not, two new masses
$m'_{\t 1} < m_{\t 1}$ and $m'_{\t 2} < m_{\t 2}$ are selected
according to the normal Breit-Wigners, repeatedly until 
$E' = \hat{m} - m'_{\t 1} - m'_{\t 2} > 0$. Then a hybrid weight 
$\tilde{G}(E)/\beta'_{\t}$ is applied, i.e.\ with the original 
$E < 0$ but the new $\beta'_{\t}$ based on $m'_{\t 1}$ 
and $m'_{\t 2}$. Overall this gives a good rendering of the desired
$\tilde{G}(E)$, Fig.~\ref{fig:threshold}\textit{b}, although a small 
discontinuity at $E = 0$ indicates that the match is not perfect. 
The $\langle E \rangle = -4.24$ GeV for the $E < 0$ part lines up 
well with the previous numbers.

The integrated cross section for the $E < 0$ part is 5.70, 6.27 and 
6.76 pb, respectively, for the three scenarios above. The first number 
clearly is too low owing to the too high $\langle x \rangle$ values.  
The difference between the last two could partly come from a further
fine-tuning of the $\hat{m}$ spectrum, but also partly from the
above-mentioned small discontinuity at $E = 0$. An intermediate value 
of 6.5 pb could be compared with the experimental signals
$8.8^{+ 1.2}_{- 1.4}$ pb for CMS and $9.0 \pm 1.3$ pb for ATLAS.
It is important to remember, however, that a realistic comparison
would need to consider the theoretical and experimental handling of
the whole threshold region. On the former side, the choice of PDF,
of $\alpha_{\mathrm{s}}$, of transition away from the Green's function
near $E = 0$, and so on, also matter. And, most interestingly, whether
higher-order effects could affect the colour singlet/octet composition
assumed here, which likely would act in favour of enhancing the
colour-singlet fraction \cite{Fadin:1990wx,Fadin:1991zw}. In the 
all-octet extreme the 6.76 pb would drop to 2.40 pb, and in the 
all-singlet one rise to 21.5 pb, with large changes also for $E >  0$.

\begin{figure}
\includegraphics[width=0.50\textwidth]{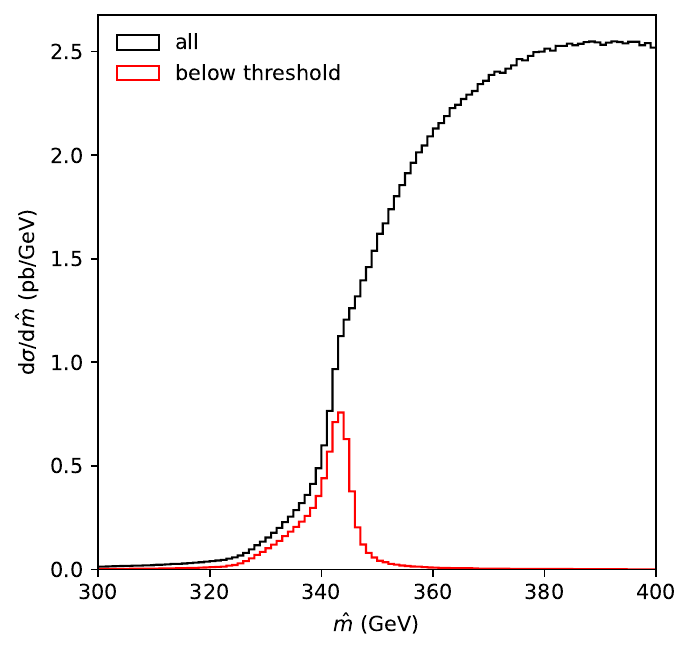}%
\includegraphics[width=0.50\textwidth]{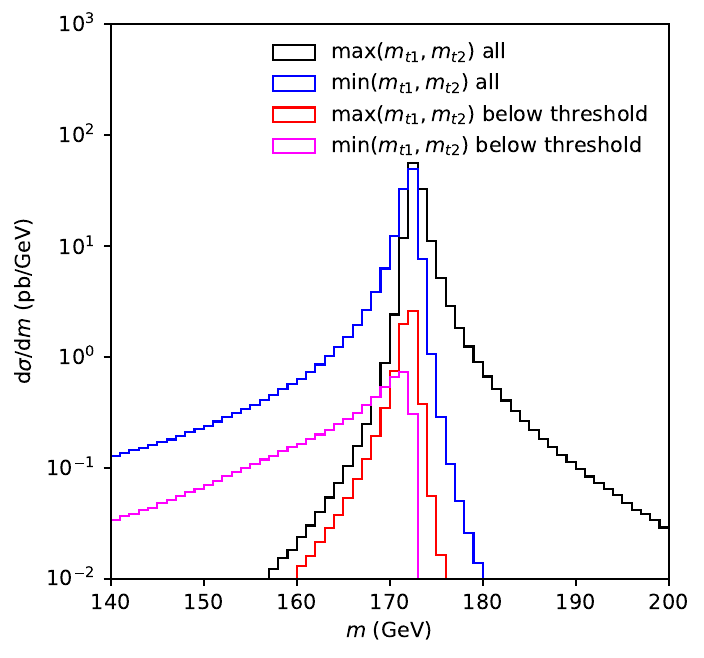}\\[-2mm]
\hspace*{0.25\textwidth}\textit{(a)}\hspace*{0.48\textwidth}\textit{(b)}
\caption{Example of event properties for the $\t\tbar$ pair mass range
$300 < \hat{m} < 400$ GeV.
\textit{(a)} Cross section $\mathrm{d}\sigma/\mathrm{d}\hat{m}$.
\textit{(b)} Cross section for the heavier and lighter of the $\t$ and
$\tbar$, respectively.}
\label{fig:prop}
\end{figure}

Examples of other event properties are shown in Fig.~\ref{fig:prop},
with BW smearing fully taken into account. The $E < 0$ contribution
dominates for low $\hat{m}$, as should be expected. But the much
larger $E > 0$ event rate means that its BW fluctuations still give
a significant contribution at low $\hat{m}$. Of note is also that the
sharp peak structure of Fig.~\ref{fig:threshold} is smeared out and
lost in the overall picture. For the individual $\t$ and $\tbar$
masses, $m_{\t 1}$ and $m_{\t 2}$ above threshold and $m'_{\t 1}$ and
$m'_{\t 2}$ below it, obviously the latter two have suppressed high-mass
tails. But the heavier of $m'_{\t 1}$ and $m'_{\t 2}$ stays close to 
on-shell, while the lighter alone can be forced well off-shell,
a familiar outcome when Breit-Wigners are involved.

In the 1989 study the below-threshold contribution was only integrated
numerically, i.e.\ no full event generation was attempted. Therefore
only the Coulomb expressions were presented in the public code. When
\textsc{Pythia}~8 \cite{Bierlich:2022pfr} was written the experimentalists
had turned to NLO codes for the $\t\tbar$, and so only the simple Born
cross sections were carried over. In the recent 8.316 release the
scenarios studied here are made available, with a set of free parameters,
such as the singlet/octet composition. Also do recall that
\textsc{Pythia} does generate complete final states starting from the
production process in focus here, with further free parameters.

There is one important catch: the top quarks are assumed unpolarized
and uncorrelated at production, and only the intermediate $\mathrm{W}$
spin is taken into account in the
$\t \to \mathrm{b}\mathrm{W}^+ \to \mathrm{b} \ell^+ \nu_{\ell}$
(or $\to \mathrm{b} \mathrm{q} \overline{\mathrm{q}}'$) decay chain. 
In the threshold region a pseudoscalar spin state should dominate, 
which thus is missed. There is room for improvement: a user hook in
the code allows you to repeatedly generate new decay angle 
configurations, assuming you have relevant matrix element code for an 
accept/reject step. Alternatively some external program could be 
delegated to handle the decays.

In summary, this note described the re-implementation of NRQCD expressions
for the $\t\tbar$ production in the threshold region, including the
contribution from states below the threshold. In spite of its
35+ years age, it may still serve as a useful reference. A new feature
is the generation of below-threshold events, which appears to work 
reasonably well. The code is made available in \textsc{Pythia}~8.316.

\acknowledgments
Valery Khoze is thanked for useful discussions, and Maria V. Garzelli
for triggering this study. This research was supported in part by
the Munich Institute for Astro-, Particle and BioPhysics (MIAPbP)
which is funded by the Deutsche Forschungsgemeinschaft (DFG, German
Research Foundation) under Germany´s Excellence Strategy
– EXC-2094 – 390783311.

\bibliographystyle{JHEP}
\bibliography{topthreshold}

\providecommand{\href}[2]{#2}\begingroup\raggedright\begin{thebibliography}{10}

\bibitem{CMS:2025kzt}
{\scshape CMS} collaboration, \emph{{Observation of a pseudoscalar excess at
  the top quark pair production threshold}},
  \href{https://doi.org/10.1088/1361-6633/adf7d3}{\emph{Rept. Prog. Phys.}
  {\bfseries 88} (2025) 087801}
  [\href{https://arxiv.org/abs/2503.22382}{{\ttfamily 2503.22382}}].

\bibitem{ATLAS:2025top}
{\scshape ATLAS} collaboration, \emph{{Observation of a cross-section
  enhancement near the ttbar production threshold in root-s = 13 TeV pp
  collisions with the ATLAS detector}},
  \href{https://arxiv.org/abs/ATLAS-CONF-2025-008}{{\ttfamily
  ATLAS-CONF-2025-008}}.

\bibitem{Fuks:2024yjj}
B.~Fuks, K.~Hagiwara, K.~Ma and Y.-J.~Zheng, \emph{{Simulating toponium
  formation signals at the LHC}},
  \href{https://doi.org/10.1140/epjc/s10052-025-13853-3}{\emph{Eur. Phys. J. C}
  {\bfseries 85} (2025) 157}
  [\href{https://arxiv.org/abs/2411.18962}{{\ttfamily 2411.18962}}].

\bibitem{Garzelli:2024uhe}
M.V.~Garzelli, G.~Limatola, S.O.~Moch, M.~Steinhauser and O.~Zenaiev,
  \emph{{Updated predictions for toponium production at the LHC}},
  \href{https://doi.org/10.1016/j.physletb.2025.139532}{\emph{Phys. Lett. B}
  {\bfseries 866} (2025) 139532}
  [\href{https://arxiv.org/abs/2412.16685}{{\ttfamily 2412.16685}}].

\bibitem{Fadin:1989fd}
V.S.~Fadin, V.A.~Khoze and T.~Sjostrand, \emph{{ON THE THRESHOLD BEHAVIOR OF
  HEAVY TOP PRODUCTION}},  in \emph{{24th Rencontres de Moriond: New Results in
  Hadronic Interactions}}, pp.~19--32, 1989.

\bibitem{Fadin:1990wx}
V.S.~Fadin, V.A.~Khoze and T.~Sjostrand, \emph{{On the Threshold Behavior of
  Heavy Top Production}}, \href{https://doi.org/10.1007/BF01614696}{\emph{Z.
  Phys. C} {\bfseries 48} (1990) 613}.

\bibitem{Fadin:1987wz}
V.S.~Fadin and V.A.~Khoze, \emph{{Threshold Behavior of Heavy Top Production in
  e+ e- Collisions}}, {\emph{JETP Lett.} {\bfseries 46} (1987) 525}.

\bibitem{Fadin:1991zw}
V.S.~Fadin and V.A.~Khoze, \emph{{Production of a pair of $t \bar{t}$ quarks
  near threshold}}, {\emph{Sov. J. Nucl. Phys.} {\bfseries 53} (1991) 692}.

\bibitem{Nason:1987xz}
P.~Nason, S.~Dawson and R.K.~Ellis, \emph{{The Total Cross-Section for the
  Production of Heavy Quarks in Hadronic Collisions}},
  \href{https://doi.org/10.1016/0550-3213(88)90422-1}{\emph{Nucl. Phys. B}
  {\bfseries 303} (1988) 607}.

\bibitem{Bierlich:2022pfr}
C.~Bierlich et~al., \emph{{A comprehensive guide to the physics and usage of
  PYTHIA 8.3}},
  \href{https://doi.org/10.21468/SciPostPhysCodeb.8}{\emph{SciPost Phys.
  Codeb.} {\bfseries 2022} (2022) 8}
  [\href{https://arxiv.org/abs/2203.11601}{{\ttfamily 2203.11601}}].

\end{thebibliography}\endgroup

\end{document}